\documentstyle[12pt]{article}
\input{epsf}
\begin{document}
{\Large Low-mass quark stars as Maclaurin spheroids}
\bigskip

{ P. Amsterdamski$^a$, T. Bulik$^b$, D. Gondek-Rosi\'nska$^{b,c}$,
W. Klu\'zniak$^{b,d}$}
\bigskip
\smallskip

$^a$
Johannes Kepler Astronomical Center, ul. Lubuska 2, 
PL-65265 Zielona G\'ora, Poland
\smallskip

$^b$
Nicolaus Copernicus Astronomical Center, ul. Bartycka 18,
 PL-00716 Warszawa, Poland
\smallskip

$^c$
 D{\'e}partement d'Astrophysique Relativiste et de Cosmologie
 UMR 8629 du CNRS, Observatoire de Paris, F-92195 Meudon  Cedex, France  
\smallskip

$^d$ 
NORDITA, 17 Blegdamsvej, Copenhagen, Denmark
\bigskip
\bigskip
\bigskip
\bigskip
\bigskip

PACS: 04.40-b,04.40.Dg, 97.10-q
\bigskip
\bigskip
\bigskip

\bigskip
Abstract
\bigskip

We show that in the limit of low masses ($M\le 0.1M_\odot$),
the orbital and rotational frequencies in
exact relativistic
numerical models of rotating, axially symmetric, quark stars coincide
with those for Maclaurin spheroids. In particular, when the eccentricity
of the spheroid
exceeds a critical value ($e>0.83458318$), circular orbits
in the equatorial plane are unstable for a range of orbital radii
outside the stellar surface. The orbital frequency in the marginally
stable orbit, and all other orbits, around Maclaurin spheroids 
goes to zero in the limit $e=1$. Maximum orbital frequencies in accreting
Galactic X-ray sources may be a poor indication of the source mass.

\bigskip
\vfill\eject
As noticed by Thomas Simpson in 1743 and by d'Alembert, 
the rotational frequency of a Maclaurin spheroid (Fig. 1)
$$\Omega^2=2\pi G\rho (1-e^2)^{1/2}e^{-3}\left\{
 (3-2e^2)\arcsin e - 3e(1-e^2)^{1/2}\right\}
\eqno (1)$$
approaches  zero not only
for $e\rightarrow0$,
a ``spheroid which departs only slightly from a sphere,''
but also for $e\rightarrow1$, a ``highly flattened spheroid'' [1].

We find
 [2] that the latter property is shared by orbital frequencies, $\omega(r)$.
Specifically,
at the equator,
$${\omega^2(a)}=  2\pi G\rho(1-e^2)^{1/2}{e^{-3} }
\left\{\arcsin e- e(1-e^2)^{1/2}\right\}.
\eqno (2)$$
In fact, 
the corresponding circular orbits (of radius $r=a$) are stable only for
eccentrities $e\le e_s\equiv0.83458318$, but
the frequency of the innermost (marginally) stable circular
orbit, present in the equatorial plane at $r>a$ 
for all $e> e_s$, also goes to zero for $e\rightarrow1$,
$${\omega_{ms}^2}=0.5276189\times 2\pi G\rho(1-e^2)^{1/2}{e^{-3} }.
\eqno (3)$$
The expressions (2), (3), for the maximum frequency in circular orbits
are plotted in Fig.~2.
  Here, $e=(1-b^2/a^2)^{1/2}$ is
the eccentricity of the spheroid, with $a$  the major and $b$ the minor
axis of the same. 
The radius of the marginally stable orbit is $r=r_{ms}=1.19820292ae\ge a$.


The above expressions are valid in  Newtonian gravity, where
the Maclaurin spheroids are bodies of uniform density
 in hydrostatic equilibrium. In general relativity there are no
incompressible bodies, but quark matter is very nearly so,
with the speed of sound close to $c/3$, one third the speed of light.

We have computed numerical models
of rotating quark stars [3] in general relativity and
have found that for stellar masses much less than that of the sun
($M<<M_\odot$), the density is nearly uniform throughout the star
and, as expected, close to the density $\rho_0$ at zero pressure
of quark matter (assumed to be self-bound). 
The stars were constrained to be axisymmetric. We have found
that for $M\le0.1M_\odot$ the Maclaurin spheroid is a very good approximation
to the quark star, at least in terms of gross stellar properties 
(Figs.~1,~2).

\begin{figure} 
\begin{center}
\leavevmode
\epsfxsize=11cm 
\epsfbox{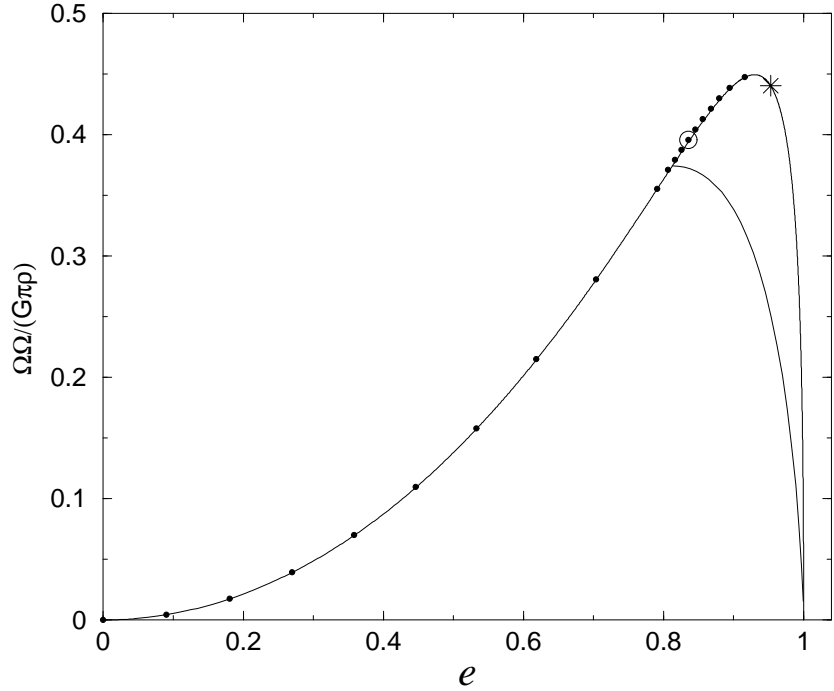}
\end{center}
\caption{The rotational frequency of Maclaurin spheroids as a
function of their eccentricity (upper curve,
the star indicates the point of onset of dynamical instability 
to a toroidal mode), and the Jacobi sequence (lower curve),
after Chandrasekhar [1]. Also shown (thick dots) are numerical models of
quark stars calculated by us with the general relativistic spectral code
of Gourgoulhon et al. [3]. The circle indicates that spheroid for
which the marginally stable orbit grazes the equator ($e=0.834583$).
}
\label{fig1}
\end{figure}
\begin{figure} 
\begin{center}
\leavevmode
\epsfxsize=11cm 
\epsfbox{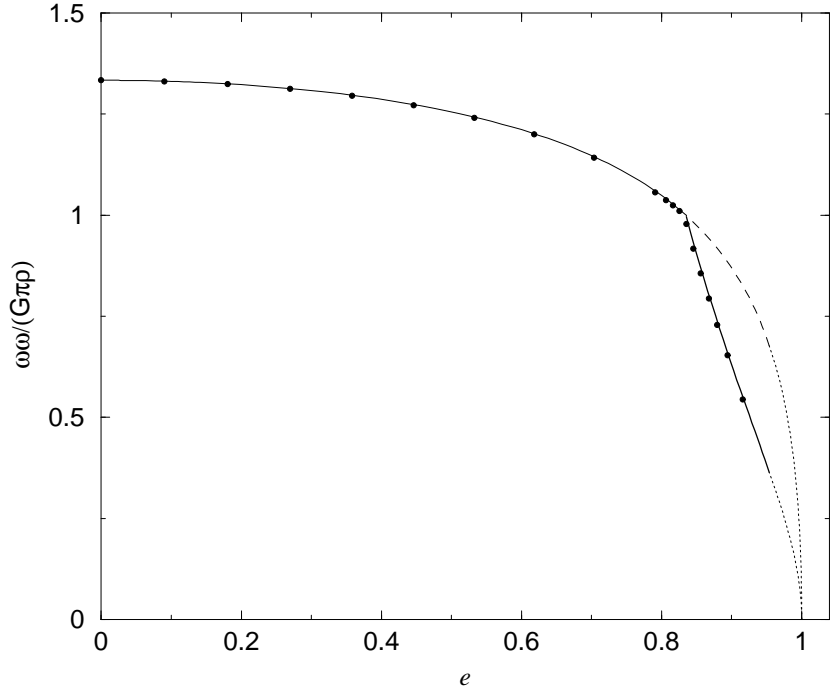}
\end{center}
\caption{The maximum frequency in stable circular orbits around the Maclaurin
spheroids as a function of the spheroid eccentricity (continuous curves).
 The thin curve is the orbital frequency at the equator, eq. (2),
the dashed portion corresponds to
unstable orbits at the equator.
 The thick curve is the orbital frequency in the
 marginally stable orbit, eq. (3). 
The dotted portions of the curves correspond to spheroid
eccentricities past the point of onset of dynamical instability, at
$e=0.95289$ [1].
Note that the curves are Newtonian,
the speed of light does not enter eqs. (2), (3). Also shown (as thick dots)
are the  numerical models of quark stars presented in Fig.~1.
}
\label{fig2}
\end{figure}
In Fig.~3 we show the results of our general-relativistic
numerical computations of the maximum orbital frequencies,
 $f=(2\pi)^{-1}\omega$, in stable circular orbits
 around a quark star of (baryon) mass $0.01M_\odot $,
modeled with the MIT-bag equation of state
$p=(\rho-\rho_0)c^2/3$. 
The actual values in Fig.~3 are for $\rho_0=4.28\times10^{14}\,$g/cm$^3$.
The results for the Dey et al.
equation of state
 [4] are similar, but the numerical
values are higher as frequencies scale with the square root of density.
The largest value of orbital frequency is obtained for
the static model ($e=0$, $\Omega=0$), and is given by the Keplerian expression
$\omega(a)=\sqrt{4\pi G\rho/3}\approx\sqrt{4\pi G\rho_0/3}$.
For every $e\le e_s$
the maximum stable orbital frequency is attained in a circular
orbit on the equator, i.e., $f= (2\pi)^{-1}\omega(a)$. 
For larger eccentricities 
$2\pi f= \omega(r_{ms})=\omega_{ms}$ and the value
of orbital frequency in the innermost stable orbit drops precipitously
as $e\rightarrow 1$. 
The numerical results
 are seen to agree with the Newtonian expressions of eqs. (2), (3).
Clearly, general-relativistic effects are
negligible for low-mass quark stars.

Because in Fig.~3 the dependence on rotational frequency 
$(2\pi)^{-1}\Omega$ is shown,
the curves are ``double valued'' (c.f. Fig.~1),
for a given rotational frequency
the larger value of orbital frequency corresponds to a spheroid/star
of lower eccentricity, the smaller to the one for larger $e$.
In this figure, the intersection of the curves for $\omega_{ms}$ and
for $\omega(a)$ on the unstable branch (dotted curve) is a mirage
caused by projection onto the $\Omega$ axis, in fact, values of about 1250 Hz
are attained for different values of $e$ in the case of
 $(2\pi)^{-1}\omega_{ms}$ and of $(2\pi)^{-1}\omega(a)$. Compare
Fig.~2, where it is clear that $\omega(a)$ and $\omega(r_{ms})$
intersect only for $e=e_s$ and $e=1$.
Dynamical instability to a toroidal deformation sets in at $e=0.95289$
in the Newtonian theory [1],
for this reason
we do not extend the curves in Fig. 3 to low orbital frequencies.

\begin{figure} 
\begin{center}
\epsfxsize=11cm 
\epsfbox{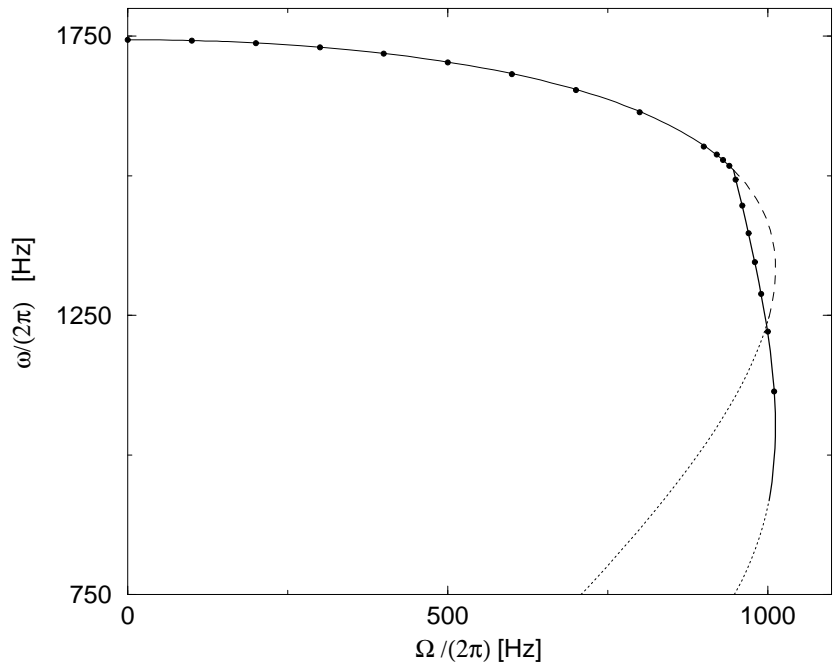}
\end{center}
\caption{The maximum orbital frequency as a function of the rotational
frequency of the Maclaurin spheroid. All symbols have
the same meaning as in Fig.~2.
}
\label{fig3}
\end{figure}

Finally, in Fig. 1 we plot the the rotational frequencies of
the numerically computed stars, superimposed on Maclaurin's 
and Jacobi's sequence [1].
The circle corresponds to the appearance of the marginally stable orbit
at $e=e_s$. Note that this occurs above the bifurcation point,
at eccentricity $e_J=0.8127$, past which the Maclaurin spheroids 
are secularly
unstable to deformation into a Jacobi ellipsoid [1].
However, this does not necessarily imply that the Newtonian
marginally stable orbit, discovered here for the Maclaurin spheroid,
is irrelevant to quark stars or other stars.
We find that in our numerical models,
the dependence of rotational frequency on eccentricity
for higher mass stars (e.g., up to $1.5M_\odot$) departs by no more
than several percent from the curve of Fig.~1, at least up to its maximum.
Similar departures are present in a 
post-Newtonian calculation of incompressible stellar models
 by Shapiro and Zane
 [5], who find that the secular instability point progressively
moves up to higher values of $e_J$ and to higher values of $\Omega$,
until the bifurcation point reaches the maximum of the rotational
frequency curve, $\Omega(e)$.

For a range of masses of compact stars, possibly existing in the real word,
 effects of general relativity may delay the onset of
secular instability to non-axisymmetric deformations past the eccentricity
where the essentially
Newtonian marginally stable orbit (related to rotation-induced
stellar flattening) appears.
What is the minimum mass of a quark star for which $e_J>e_s$?
 Calculations in ref. [5]  indicate that
the bifurcation point occurs at
$e_J=0.835$ for $GM/(Rc^2)=0.013 $, where $R$ is the radius
of the non-rotating star of gravitational mass $M$.
 For the quark stars presented here,
this would occur already for a star with $M=0.03M_\odot$,
for which $R=3.2\,$km. It would appear then,
that in the mass range $\sim(0.03 M_\odot, 0.1 M_\odot)$
quark stars are well approximated by the Maclaurin spheroids
{\sl and} post-Newtonian
effects may stabilize the quark star at eccentricities sufficiently high
that the purely Newtonian marginally stable orbit is present outside
the stellar surface, for $e>e_s$.

The expressions (2), (3), for the orbital frequencies around Maclaurin
spheroids, and the presence of the marginally stable orbit [6] in their
Newtonian theory are interesting in their own right.
In addition, the gravity of Maclaurin spheroids provides a consistent
Newtonian framework in which the influence of the marginally stable
orbit on accretion flow can be analytically investigated in two or
three dimensions---up to now analytic work on accretion in
black hole and neutron-star systems often relied on the convenient,
but artificial, one dimensional ``potentials'' of the
Paczy\'nski-Wiita type [7].

We note that the kHz QPO phenomenon (millisecond variability
of X-ray flux in bright Galactic sources [8]) has been interpreted
as giving support to the presence of the general-relativistic marginally
stable orbit in accreting neutron stars, and an application of the
relativistic equivalent of Kepler's third law to
the specific maximum values
$\approx1\,$kHz of the observed frequencies in several such systems
yields stellar masses $\approx2M_\odot$ in that interpretation [9].
But we saw that the same value of $\omega_{ms}/(2\pi)\approx 1\,$kHz
can be attained in a (Newtonian) marginally stable orbit for a 
rapidly rotating star
of rather low mass (e.g., $10^{-2}M_\odot$ or $10^{-1}M_\odot$).
Essentially, this is because in general relativity
the gravitational radius $r_g=2GM/c^2$ is a natural unit of length,
and the characteristic
frequency is $\sqrt{GM/r_g^3}\propto M^{-1}$, requiring a large mass for
a low frequency, while for the Maclaurin spheroid
the characteristic unit of length is $ae$ and the corresponding 
characteristic frequency is  $\omega_c=\sqrt{GM/(ae)^3}$.
 Since by mass conservation $a^3(1-e^2)^{1/2}=\,$ const,
 $\omega_c^2\propto e^{-3}(1-e^2)^{1/2}$ for the Maclaurin spheroids
and  $\omega_c$    vanishes for $e=1$.

We conclude that it would be premature to assign a mass value to
a compact star based on observations of maximal orbital frequency
alone, even if there is no doubt that the frequency is obtained in
the marginally stable orbit.
In a
source where much of the luminosity is released in an accretion disk,
it would be difficult in most cases to discriminate directly
between a star of mass $1.4M_\odot$ with a  $10\,$km radius
 and, say,  a $0.1M_\odot$ star of radius $5\,$km.
One indication of a low mass could be a lower photon flux during an
X-ray burst, when the luminosity is thought to reach the Eddington limit value
$L_{\rm Edd}=9\times10^{37}(M/M_\odot)\,$erg/s~[10].
 It could well be that many of the kHz QPO
sources are in fact rapidly rotating low-mass quark stars.
Fig. 3 makes it clear that in such a case the maximum orbital
frequency may have a value far lower than would be the case
for a spherical star.

Low-mass quark stars are well described by
Maclaurin spheroids. This is the only known example of an accurate
analytic description of a compact stellar remnant
composed of matter at supranuclear density.
We expect that in general,
qualitative features of the exterior gravitational
potential of Maclaurin spheroids may be relevant
to a discussion of rapidly rotating stars.
White dwarfs do not have uniform density, but if
they rotate very rapidly, as they are expected to in cataclysmic variables,
their orbital frequencies may turn out to be closer to those of the Maclaurin
spheroids than to those of a point mass. 
Neutron stars are unstable below $\sim0.1M_\odot$ while the results
presented here are strictly applicable 
only to low-mass compact stars, but we speculate
that the exterior metric of massive ($M\sim M_\odot$) neutron stars
and quark stars in general relativity may share some features with
the behavior uncovered here for Maclaurin spheroids. For example, 
in black holes frame dragging draws the marginally stable orbit
 in towards the axis of rotation,
as the black hole spin increases, but
it is known
that for rapidly rotating massive quark stars
rotation acts in the opposite sense: the marginally stable orbit
is pushed out [11].
Hints of the same effect are observed in neutron star models at highest masses.
We believe that the origin of this behavior is Newtonian, and 
point out that
effects of stellar flattening, present already
in  Newtonian theory, counteract
relativistic effects of frame dragging.

\vfill\eject

\parindent=0pt
REFERENCES AND NOTES

[1] Chandrasekhar, S., {\it Ellipsoidal Figures of Equilibrium}
(Yale University Press, New Haven 1969).

[2] Amsterdamski, P., Klu\'zniak, W., in preparation (2001).

[3] Numerical models of static quark stars in general relativity
have been constructed by Itoh, N., Progr. Theor. Phys. 44, 291 (1970);
 Brecher, K., and Caporaso, G., Nature 259, 377 (1976);
Witten, E., Phys. Rev. D 30, 272 (1984);
Alcock, Ch., Farhi, E., and Olinto, A., ApJ 310, 261 (1986);
and others. The first accurate, fully relativistic
calculations
of rotating quark stars were published by
 Gourgoulhon, E. et al., Astron. Astrophys. 349, 851 (1999);
Stergioulas,  N., Klu\'zniak, W., and Bulik, T.,
 Astron. Astrophys. 352, L116 (1999).
 We use Gourgoulhon's code here.

[4] Dey, M., et al., Physics Letters B, {438}, 123 (1998).

[5] Shapiro, S.L., and Zane, S., ApJ 117, 531 (1998).

[6] 
The effect on orbital frequency
of quadrupole and octupole moments of mass distribution has
been  pointed out  by 
Sibgatulin, N.R., and Sunyaev R.A., Astron. Lett. 24, 774 (1998);
and Shibata, M., and Sasaki, M., Phys. Rev. D60 084002 (1999);
 but the Newtonian
origin of this phenomenon apparently went unrecognized.
The marginally stable orbit in low-mass quark stars was first discovered
numerically (Klu\'zniak, W., Bulik, T., and Gondek-Rosi\'nska
{\it The 4th INTEGRAL Workshop} in press, astro-ph/0011517,
 and the Newtonian character
of the marginally stable orbit there was stressed
by Zdunik, L., and Gourgoulhon, E., astro-ph/0011028 (2000).

[7] Paczy\'nski, B., and Wiita, P.J., Astron. Astrophys. 88, 23 (1980);
Kato, S., Mineshige, S., and Fukue, J. {\it Black Hole Accretion Disks}
(Kyoto University Press, Kyoto 1998).

[8] van der Klis, M., Ann. Rev. Astron. Astrophys. 38,  717 (2000).

[9] Klu\'zniak, W., Michelson, P., Wagoner, R. V., ApJ, 358, 538 (1990);
Kaaret, P., Ford, E. C., and Chen, K., ApJ, 480, 127 (1997);
Zhang, W., Strohmayer, T. E., and Swank, J. H. 1997, ApJ, 482, L167;
Klu\'zniak, W., ApJ, 509, L37 (1998).

[10] Margon, B., and Ostriker, P., ApJ 186, 91 (1973).

[11] Stergioulas, N., Klu\'zniak, W., and Bulik, T.,
 Astron. Astrophys. 352, L116 (1999).

W.K. gratefully acknowledges the hospitality of NORDITA, where some of the
analytic calculations were completed.
Research supported in part by KBN grant 2P03D 00418.

\end{document}